\documentclass[twocolumn,showpacs,preprintnumbers,amsmath,amssymb]{revtex4}

\usepackage{graphicx}
\usepackage{dcolumn}
\usepackage{bm}

\def\gsim{\mathop {\vtop {\ialign {##\crcr 
$\hfil \displaystyle {>}\hfil $\crcr \noalign {\kern1pt \nointerlineskip } 
$\,\sim$ \crcr \noalign {\kern1pt}}}}\limits}
\def\lsim{\mathop {\vtop {\ialign {##\crcr 
$\hfil \displaystyle {<}\hfil $\crcr \noalign {\kern1pt \nointerlineskip } 
$\,\,\sim$ \crcr \noalign {\kern1pt}}}}\limits}

\begin{document}

\preprint{APS/123-QED}

\title{
Quantum Valence Criticality as Origin of Unconventional Critical Phenomena
}

\author{Shinji Watanabe}
\author{Kazumasa Miyake}%
\affiliation{%
Division of Materials Physics, Department of Materials Engineering Science, Graduate School of
Engineering Science, Osaka University, Toyonaka, Osaka 560-8531, Japan
}%

\date{April 6, 2010}

\begin{abstract}
It is shown that unconventional critical phenomena commonly observed in paramagnetic metals YbRh$_2$Si$_2$, 
YbRh$_2$(Si$_{0.95}$Ge$_{0.05}$)$_2$, and $\beta$-YbAlB$_4$ is naturally explained 
by the quantum criticality of Yb-valence fluctuations.  
We construct the mode coupling theory taking account of local correlation effects of f electrons and find that unconventional criticality is caused by the locality of the valence fluctuation mode. 
We show that measured low-temperature anomalies such as divergence of uniform spin susceptibility $\chi\sim T^{-\zeta}$ with $\zeta\sim 0.6$    
giving rise to a huge enhancement of the Wilson ratio and the emergence of 
$T$-linear resistivity are explained in a unified way.  
\end{abstract}

\pacs{71.27.+a, 75.20.Hr, 71.10.-w, 71.20.Eh}
\maketitle

Quantum critical phenomena have been one of the central issues in condensed matter physics 
for the past two decades. 
The nature of the quantum critical point (QCP) emerging when the magnetically-ordered temperature 
is suppressed to absolute zero has been intensively studied 
and the role of spin fluctuations in the critical phenomena has been well understood~\cite{Moriya,MT,Hertz,Millis}. 

However, anomalous critical phenomena, which do not follow the conventional spin-fluctuation 
theories~\cite{Moriya,MT,Hertz,Millis} have been discovered in 
paramagnetic metal phases in YbRh$_2$Si$_2$~\cite{Trovarelli,Gegenwart2007},
YbRh$_2$(Si$_{0.95}$Ge$_{0.05}$)$_2$~\cite{Gegenwart2005,Custers}, and 
$\beta$-YbAlB$_4$~\cite{Nakatsuji}. 
The most striking anomaly is that low-temperature 
uniform spin susceptibility exhibits divergent behavior $\chi(T)\sim T^{-\zeta}$ 
with the anomalous critical exponent $\zeta=0.6$ in YbRh$_2$(Si$_{0.95}$Ge$_{0.05}$)$_2$ 
and $\zeta=0.5$ in $\beta$-YbAlB$_4$ 
in spite of no sign of a ferromagnetic phase nearby. 
In these materials, the 
Sommerfeld constant exhibits the logarithmic divergence 
$\gamma_{\rm e}=C_{\rm e}/T\sim -{\rm ln}T$~\cite{Trovarelli,Custers,Nakatsuji}, giving rise to a 
large Wilson ratio, i.e., a dimensionless ratio of $\chi$ to $\gamma_{\rm e}$, $R_{\rm W}=17.5$ in 
YbRh$_2$(Si$_{0.95}$Ge$_{0.05}$)$_2$~\cite{Gegenwart2005}, 
and $R_{\rm W}=6.5$ in $\beta$-YbAlB$_4$~\cite{Nakatsuji}, 
exceeding the conventional strong-coupling value $R_{\rm W}=2$. 
The linear-$T$ dependence of low-$T$ resistivity emerges in a wide-$T$ range 
in YbRh$_2$Si$_2$~\cite{Trovarelli}, YbRh$_2$(Si$_{0.95}$Ge$_{0.05}$)$_2$~\cite{Custers}, 
and also $\beta$-YbAlB$_4$~\cite{Nakatsuji}. 
These observations suggest that there exists a new class of materials showing similar critical phenomena, which are unconventional.

So far, 
to explain YbRh$_2$Si$_2$ and YbRh$_2$(Si$_{0.95}$Ge$_{0.05}$)$_2$, theoretical efforts have been made~\cite{Si,Coleman,Misawa}. In particular, a scenario asserting that f electrons undergo a localized to itinerant transition was extensively discussed~\cite{Si,Coleman}. 
However, YbRh$_2$Si$_2$ shows a large Sommerfeld constant as $\sim$ 1.7/JmolK$^2$ 
even inside of the antiferromagnetic (AF) phase~\cite{Custers}. 
This fact indicates that 
heavy quasiparticles are responsible for the AF state. 
Indeed, a band-structure calculation showed evidence contrary to the scenario by demonstrating that 
a tiny valence change of Yb can explain the Hall-coefficient measurement in YbRh$_2$Si$_2$~\cite{Paschen}, 
which seems to be a basis of the above scenario~\cite{Norman}.

Recently, in $\beta$-YbAlB$_{4}$ the valence of Yb has been detected as 
Yb$^{+2.75}$ (0.75 4f hole per Yb) at $T=20$~K, 
suggesting strong valence fluctuations~\cite{Ohkawa}. 
Close relation of anomalous critical phenomena and valence fluctuations 
has been also indicated in 
Ce$_{0.9-x}$Th$_{0.1}$La$_{x}$~\cite{Lashley} and in YbAuCu$_4$~\cite{Wada}. 
Both are related to typical valence-transition materials: One is 
Ce metal, well known as $\gamma$-$\alpha$ transition~\cite{Cevalence}, and 
the other is YbInCu$_4$~\cite{Felner}, both showing a discontinuous valence jump 
of a Ce and Yb ion, respectively, when $T$ and $P$ are varied. 
Since the first-order valence transition is an isostructural transition, 
the critical end point exists in the $T$-$P$ (and -chemical composition) phase diagram, 
as in the liquid-gas transition. 

At $x\approx 0.1$ in Ce$_{0.9-x}$Th$_{0.1}$La$_{x}$, at which the critical end point is most suppressed 
and is close to $T=0$~K~\cite{Lashley}, critical phenomena rising from 
the quantum critical end point were revealed: $T$-linear resistivity 
emerges prominently and uniform spin susceptibility is enhanced at low $T$ 
giving rise to a large Wilson ratio, $R_{\rm W}\sim 3$. 
In YbAuCu$_4$, the uniform spin susceptibility is enhanced as 
$\chi(T)\sim T^{-0.6}$ as $T$ decreases 
in spite of the fact that the AF transition takes place at $T_{\rm N}=0.8$~K~\cite{Sarrao}, 
similarly to YbRh$_2$Si$_2$. 
Furthermore, 
the sharp Yb-valence crossover temperature $T^{*}_{\rm v}(H)$ is induced 
by applying a magnetic field~\cite{Wada}, suggesting that YbAuCu$_4$ is located 
in the vicinity of the quantum critical end point of the valence transition. 
It should also be noted that 
the $T$-$H$ phase diagram of YbAuCu$_4$~\cite{Wada} closely resembles 
that of YbRh$_2$Si$_2$~\cite{Gegenwart2007}. The $H$ dependence of the crossover temperature $T^{*}(H)$, 
emerging in several physical quantities~\cite{{Gegenwart2007}}, whose origin is unclear 
in YbRh$_2$Si$_2$ is quite similar to the $T^{*}_{\rm v}(H)$ in YbAuCu$_4$. 

These observations strongly suggest the importance of quantum criticality of valence transition 
as a key mechanism of unconventional critical phenomena. 
From this viewpoint, 
in this Letter, we resolve this outstanding puzzle by showing that 
(a) uniform spin susceptibility diverges with anomalous criticality 
$\chi(T)\sim T^{-\zeta}$ with 
$0.5 \lsim \zeta \lsim 0.7$ 
in paramagnetic metals even without proximity to a ferromagnetic phase and 
(b) $T$-linear resistivity emerges in the wide-$T$ range.

Let us start our discussion by introducing a minimal model 
which describes the essential part of the Ce- and Yb-based systems 
in the standard notation: 
\begin{equation}
H=H_{\rm c}+H_{\rm f}+H_{\rm hyb}+H_{U_{\rm fc}}, 
\label{eq:PAM} 
\end{equation}
where 
$H_{\rm c}=\sum_{{\bf k}\sigma}\varepsilon_{\bf k}
c_{{\bf k}\sigma}^{\dagger}c_{{\bf k}\sigma}$, 
$H_{\rm f}=\varepsilon_{ \rm f}\sum_{i\sigma}n^{ \rm f}_{i\sigma}
+U_{\rm ff}\sum_{i=1}^{N}n_{i\uparrow}^{ \rm f}n_{i\downarrow}^{ \rm f}
$, 
$H_{\rm hyb}=\sum_{{\bf k}\sigma}V_{\bf k}\left(
f_{{\bf k}\sigma}^{\dagger}c_{{\bf k}\sigma}+c_{{\bf k}\sigma}^{\dagger}f_{{\bf k}\sigma}
\right)$, 
and 
$
H_{U_{\rm fc}}=
U_{\rm fc}\sum_{i=1}^{N}n_{i}^{ \rm f}n_{i}^{ c}
$. 
The $U_{\rm fc}$ term is the Coulomb repulsion between 4f and conduction electrons (holes) 
in Ce (Yb) systems, 
which is considered to play an important role in the valence transition~\cite{M07}. 

To consider correlation effects by $U_{\rm ff}$, we employ the slave-boson large-$N$ expansion 
framework~\cite{OM}. The Hamiltonian Eq.~(\ref{eq:PAM}) is generalized to the case with $N$-fold degeneracy 
from $\sigma=\uparrow, \downarrow$, and the slave-boson operator $b_{i}$ is introduced to eliminate 
the doubly-occupied state for $U_{\rm ff}\to\infty$ 
under the constraint $\sum_{m}n^{\rm f}_{im}+Nb^{\dagger}_{i}b_{i}=1$. 
The Lagrangian is written as 
$
{\cal L}={\cal L}_{0}+{\cal L}'$: 
\begin{eqnarray}
&{\cal L}_{0}&=\sum_{{\bf k}m}c^{\dagger}_{{\bf k}m}
\left(
\partial_{\tau}+\bar{\varepsilon}_{\bf k}
\right)
c_{{\bf k}m}
+\sum_{{\bf k}{\bf k'}m}f^{\dagger}_{{\bf k}m}
\left(
\partial_{\tau}+\bar{\varepsilon}^{\rm f}_{{\bf k}-{\bf k'}}
\right)
f_{{\bf k'}m}
\nonumber
\\
&+&\frac{V}{\sqrt{N_{\rm s}}}\sum_{{\bf k}{\bf k'}m}
\left(
c^{\dagger}_{{\bf k}m}f_{{\bf k'}m}b^{\dagger}_{{\bf k}-{\bf k'}}
+{\rm h}.{\rm c}.
\right)
+\frac{N}{N_{\rm s}}\sum_{{\bf k}{\bf k'}}
b^{\dagger}_{\bf k}\lambda_{{\bf k}-{\bf k'}}b_{\bf k'}
\nonumber
\\
&{\cal L'}&=
-\frac{U_{\rm fc}}{2}\sum_{im}
\left(
n^{\rm c}_{im}+n^{\rm f}_{im}
\right)
+\frac{U_{\rm fc}}{N}\sum_{imm'}
n^{\rm f}_{im}n^{\rm c}_{im'}, 
\nonumber
\end{eqnarray}
where 
$\lambda_{\bf k}$ is the Lagrange multiplier to impose the constraint, 
and 
$
\bar{\varepsilon}_{\bf k}\equiv\varepsilon_{\bf k}
+\frac{U_{\rm fc}}{2}
$ 
and 
$
\bar{\varepsilon}^{\rm f}_{{\bf k}-{\bf k'}}\equiv
\left(
\varepsilon_{\rm f}+\frac{U_{\rm fc}}{2}
\right)
\delta_{{\bf k}{\bf k'}}
+\frac{1}{\sqrt{N_{\rm s}}}
\lambda_{{\bf k}-{\bf k'}}
$. 
We here separate $\cal L$ as ${\cal L}_{0}$ and ${\cal L'}$ 
to perform the expansion with respect to the $U_{\rm fc}$ term 
after taking account of the local correlation of the $U_{\rm ff}$ term. 

For $\exp(-S_{0})$ with 
the action 
$
S_{0}=\int_{0}^{\beta}d\tau{\cal L}_{0}(\tau)
$, 
the saddle point solution 
is obtained via the stationary condition $\delta S_{0}=0$ 
by approximating spatially uniform 
and time independent ones, i.e., $\lambda_{\bf q}=\lambda\delta_{\bf q}$ 
and $b_{\bf q}=b\delta_{\bf q}$. 
The solution is obtained by solving 
mean-field equations $\partial S_{0}/\partial\lambda=0$ 
and $\partial S_{0}/\partial b=0$ self-consistently.

For 
$
S'=\int_{0}^{\beta}d\tau{\cal L'}(\tau), 
$ 
we introduce the identity applied by 
a Stratonovich-Hubbard transformation
$
{\rm e}^{-S'}
=\int{\cal D}\varphi
\exp[
\sum_{im}
\int_{0}^{\beta}d\tau
\{
-\frac{U_{\rm fc}}{2}\varphi_{im}(\tau)^2
+i\frac{U_{\rm fc}}{\sqrt{N}}
(
c_{im}f^{\dagger}_{im}-f_{im}c^{\dagger}_{im}
)
\}
\varphi_{im}(\tau)
)
].
$ 
The partition function is expressed as 
$
Z=\int{\cal D}(cc^{\dagger}ff^{\dagger}\varphi)\exp(-S)
$ 
with $S=S_{0}+S'$. 
By performing Grassmann number integrals for $cc^{\dagger}$ 
and $ff^{\dagger}$, 
we obtain 
$Z=\int{\cal D}\varphi\exp(-S[\varphi])$ 
with 
$
S[\varphi]=
\sum_{m}\sum_{\bar{q}}\varphi_{m}(\bar{q})\varphi_{m}(-\bar{q})
-{\rm Tr}{\rm ln}[-\hat{G}_{0}^{-1}+\hat{V}]
-\frac{\beta N}{\sqrt{N_{\rm s}}}\lambda|b|^{2}
+\beta Nq_{0}\sqrt{N_{\rm s}}\lambda. 
$
Here, the abbreviation $\bar{q}\equiv({\bf q},i\omega_{l})$ with $\omega_{l}=2l\pi T$ is used, and 
$\hat{G}_{0}$ and $\hat{V}$ are defined as
\[ 
\hat{G}_{0}^{-1}\equiv
\left[ \begin{array}{cc}
i\varepsilon_{n}-\bar{\varepsilon}_{\bf k} & -\frac{Vb^{*}}{\sqrt{N_{\rm s}}} \\
-\frac{Vb}{\sqrt{N_{\rm s}}} & i\varepsilon_{n}-\bar{\varepsilon}^{\rm f}_{\bf 0}
\end{array}
\right]\delta_{{\bf k}{\bf k'}}, 
\hat{V}\equiv
\left[ \begin{array}{cc}
0 & \tilde{\varphi}_{m\bar{k}\bar{k}'} \\
\tilde{\varphi}_{m\bar{k}\bar{k}'} & 0
\end{array}
\right] 
\] 
with $\tilde{\varphi}_{m\bar{k}\bar{k}'}\equiv 
\frac{U_{\rm fc}}{\sqrt{\beta NN_{\rm s}}}\varphi_{m}(\bar{k}-\bar{k}')$, 
respectively. 
Here, 
$
\bar{k}\equiv({\bf k},i\varepsilon_{n})
$ 
with
$\varepsilon_{n}=(2n+1)\pi T$, 
and 
each matrix element of $\hat{G}_{0}$ is defined as 
$(\hat{G}_{0})_{11}\equiv G^{\rm cc}_{0}$, 
$(\hat{G}_{0})_{12}\equiv G^{\rm cf}_{0}$, 
$(\hat{G}_{0})_{21}\equiv G^{\rm fc}_{0}$, and
$(\hat{G}_{0})_{22}\equiv G^{\rm ff}_{0}$. 
By using 
$
{\rm Tr}{\rm ln}[-\hat{G}_{0}^{-1}+\hat{V}]
={\rm Tr}{\rm ln}[-\hat{G}_{0}^{-1}]
-\sum_{n=1}^{\infty}\frac{1}{n}{\rm Tr}(\hat{G}_{0}\hat{V})^{n}, 
$
we obtain 
\begin{eqnarray}
&S\left[\varphi\right]&=\sum_{m}\left[
\frac{1}{2}
\sum_{\bar{q}}\Omega_{2}(\bar{q})
\varphi_{m}(\bar{q})\varphi_{m}(-\bar{q})
\right.
\nonumber
\\
&+&
\left.
\sum_{\bar{q}_1,\bar{q}_2,\bar{q}_3}
\Omega_{3}(\bar{q}_1,\bar{q}_2,\bar{q}_3)
\varphi_{m}(\bar{q}_1)
\varphi_{m}(\bar{q}_2)
\varphi_{m}(\bar{q}_3)
\delta\left(\sum_{i=1}^{3}\bar{q}_{i}\right)
\right.
\nonumber \\
&+&
\left. 
\sum_{\bar{q}_1,\bar{q}_2,\bar{q}_3,\bar{q}_4}
\Omega_{4}(\bar{q}_1,\bar{q}_2,\bar{q}_3,\bar{q}_4)
\right.
\nonumber 
\\
&{\times}&
\left.
\varphi_{m}(\bar{q}_1)
\varphi_{m}(\bar{q}_2)
\varphi_{m}(\bar{q}_3)
\varphi_{m}(\bar{q}_4)
\delta\left(\sum_{i=1}^{4}\bar{q}_{i}\right)
+\cdots
\right]. 
\label{eq:Fexpand}
\end{eqnarray}
Here, 
constant terms independent of $\varphi_{m}(\bar{q})$ are omitted in Eq.~(\ref{eq:Fexpand}) 
since they merely shift the origin of the free energy of the system. 
The coefficient of the quadratic term is given by 
\begin{eqnarray}
\Omega_{2}({\bf q},i\omega_{l})=U_{\rm fc}
\left[1-\frac{2U_{\rm fc}}{N}
\left\{
\chi^{\rm ffcc}_{0}({\bf q},i\omega_{l})-
\chi^{\rm cfcf}_{0}({\bf q},i\omega_{l})
\right\}
\right], 
\label{eq:v2}
\end{eqnarray}
where $\chi^{\alpha\beta\gamma\delta}_{0}({\bf q},i\omega_{l})
\equiv -\frac{T}{N_{\rm s}}\sum_{{\bf k},n}
G_{0}^{\alpha\beta}({\bf k}+{\bf q},i\varepsilon_{n}+i\omega_{l})
G_{0}^{\gamma\delta}({\bf k},i\varepsilon_{n})
$. 
Since long wave length $|{\bf q}|\ll q_{\rm c}$ around ${\bf q}={\bf 0}$ 
and low frequency $|{\omega}|\ll \omega_{\rm c}$ regions play dominant roles 
in critical phenomena with $q_{\rm c}$ and $\omega_{\rm c}$ being cutoffs 
for momentum and frequency in the order of inverse of the lattice constant 
and the effective Fermi energy, respectively, 
$\Omega_{i}$ for $i=2,3$, and 4 are expanded for $q$ and $\omega$ around 
$({\bf 0},0)$: 
\begin{eqnarray}
\Omega_{2}({\bf q},i\omega_{l})
\approx 
\eta
+Aq^{2}+C\frac{\left|\omega_{l}\right|}{q}
\label{eq:O2expand}
\end{eqnarray}
where 
$
\eta=U_{\rm fc}
[1-\frac{2U_{\rm fc}}{N}
\{
\chi^{\rm ffcc}_{0}({\bf 0},0)-
\chi^{\rm cfcf}_{0}({\bf 0},0)
\}
]
$, 
$\Omega_{3}(q_1,q_2,q_3)\approx v_3/\sqrt{\beta N_{\rm s}}$, and 
$\Omega_{4}(q_1,q_2,q_3,q_4)\approx v_4/(\beta N_{\rm s})$. 

\begin{figure}[h]
\includegraphics[width=70mm]{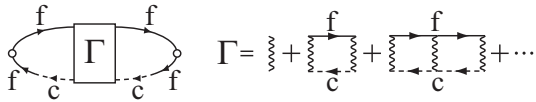} 
\caption{\label{fig:RPA} 
Feynman diagrams for dynamical valence susceptibility and 
dynamical spin susceptibility for f electrons. 
Solid lines and dashed lines represent the f- and conduction-electron 
Green functions, $G^{\rm ff}_{0}$ and $G^{\rm cc}_{0}$, respectively.
Wiggly lines represent $U_{\rm fc}$. 
}
\end{figure}

Different from ordinary critical phenomena of spin fluctuations~\cite{Moriya,MT,Hertz,Millis}, 
there appears a cubic term in Eq.~(\ref{eq:Fexpand}) in general for the 
valence fluctuation case~\cite{cubic}. 
Let us here apply 
the Hertz's renormalization-group procedure~\cite{Hertz} to $S[\varphi]$: 
(a) Integrating out high momentum and frequency parts for $q_{\rm c}/s<q<q_{\rm c}$ and 
$\omega_{\rm c}/s^{z}<\omega<\omega_{\rm c}$, respectively, with 
$s$ being a dimensionless scaling parameter $(s\ge 1)$ and $z$ the dynamical exponent. 
(b) Scaling of $q$ and $\omega$ by $q'=sq$ and $\omega'=s^{z}\omega$. 
(c) Re-scaling of $\varphi$ by $\varphi'({\bf q'},\omega')=s^{a}\varphi({\bf q'}/s,\omega'/s)$. 
Then, we dertermined that to make the Gaussian term in Eq.~(\ref{eq:Fexpand}) scale invariant, 
$a$ must satisfy $a=-(d+z+2)/2$ with $d$ spatial dimension and the dynamical exponent $z=3$. 
The renormalization-group equations for coupling constants $v_{j}$ are derived as 
$
\frac{dv_{3}}{ds}=\left[6-(d+z)\right]v_{3}+O(v_{3}^2), 
$
and 
$
\frac{dv_{4}}{ds}=\left[4-(d+z)\right]v_{4}+O(v_{4}^2),
$
for cubic and quadratic terms, respectively. 
By solving these equations, it is shown that higher order terms than 
the Gaussian term are irrelevant 
\begin{eqnarray}
\lim_{s\to\infty}v_{j}(s)=0 \ \ {\rm for} \ \ j\ge 3
\label{eq:vterms}
\end{eqnarray}
for $d+z>6$. For the case of $d=3$ and $z=3$, it is shown that the cubic term is 
marginally irrelevant~\cite{M07}. 
Hence, the universality class of the criticality of valence fluctuations belongs to 
the Gaussian fixed point. 
This implies that critical valence fluctuations are 
qualitatively described by the RPA framework with respect to $U_{\rm fc}$. 
The coefficient of the Gaussian term in Eq.~(\ref{eq:Fexpand}) is nothing but the 
inverse of the valence susceptibility 
$\Omega_{2}({\bf q},i\omega_{l})\equiv\chi_{\rm v}({\bf q},i\omega_{l})^{-1}$. 
Since evaluation of $\chi^{\rm ffcc}_{0}({\bf q},i\omega_{l})$ and 
$\chi^{\rm cfcf}_{0}({\bf q},i\omega_{l})$ using the saddle point solution for $\exp(-S_{0})$ 
concludes $\chi^{\rm ffcc}_{0}\gg \chi^{\rm cfcf}_{0}$ 
(see Fig.~2 and text below), it turns out that 
$\chi_{\rm v}$ is expressed by the RPA form 
$\chi_{\rm v}({\bf q},i\omega_{l})
=\int_{0}^{\beta}d\tau\langle T_{\tau}
n_{\rm f}({\bf q},\tau)n_{\rm f}({\bf -q},0)
\rangle
{\rm e}^{i\omega_{l}\tau} 
\approx U_{\rm fc}^{-1}[1-\frac{2U_{\rm fc}}{N}\chi^{\rm ffcc}_{0}({\bf q},i\omega_{l})]^{-1}$, 
as shown in Fig.~\ref{fig:RPA}. 

An important consequence of this result is that 
dynamical f-spin susceptibility 
$
\chi_{\rm f}^{+-}({\bf q},i\omega_{l})
\equiv \int_{0}^{\beta}d\tau\langle T_{\tau}
S_{\rm f}^{+}({\bf q},\tau)S_{\rm f}^{-}({\bf -q},0)\rangle
{\rm e}^{i\omega_{l}\tau}
$
has a common structure with $\chi_{\rm v}$ 
in the RPA framework as shown in Fig.~\ref{fig:RPA}. 
At the quantum critical end point of the valence transition, namely, the QCP, 
the valence susceptibility 
$\chi_{\rm v}({\bf 0},0)$ diverges. The common structure indicates that 
$\chi_{\rm f}^{+-}({\bf 0},0)$ 
also diverges at the QCP. 
The uniform spin susceptibility is given by $\chi\approx\chi^{\rm f}_{\rm s}\approx\frac{3}{2}\mu_{\rm B}^{2}g_{\rm f}^{2}\chi^{+-}_{\rm f}({\bf 0},0)$ with 
$\chi^{\rm f}_{\rm s}$ uniform f-spin susceptibility, 
$\mu_{\rm B}$ the Bohr magneton, and $g_{\rm f}$ Lande's g factor for f electrons. 
This gives a qualitative explanation for the fact that 
uniform spin susceptibility diverges at the QCP of valence transition 
under a magnetic field, 
which was shown by the slave-boson mean-field theory applied to Eq.~(\ref{eq:PAM})~\cite{WTMF}. 
Numerical calculations for Eq.~(\ref{eq:PAM}) 
in $d=1$ by the DMRG~\cite{WTMF} and in $d=\infty$ by the DMFT~\cite{sugibayashi} 
also showed the simultaneous divergence of $\chi_{\rm v}$ and uniform spin susceptibility 
under the magnetic field, reinforcing the above argument based on RPA.

\begin{figure}
\includegraphics[width=70mm]{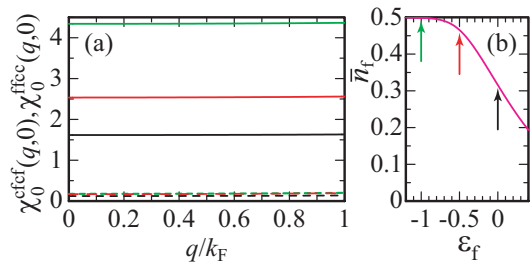}
\caption{\label{fig:ffcc} (color) 
(a) $q$ dependence of 
$\chi^{\rm ffcc}_{0}(q,0)$ (solid line) and $\chi^{\rm cfcf}_{0}(q,0)$ (dashed line)  
calculated by using saddle point solution of $\exp(-S_{0})$ 
for $\varepsilon_{\rm f}=-1$ (green), $-0.5$ (red), and $0.0$ (black). 
(b) $\bar{n}_{\rm f}$ vs $\varepsilon_{\rm f}$. 
(a) and (b) are results for $D=1$, $V=0.5$, and $U_{\rm ff}=\infty$ at $n=7/8$. 
}
\end{figure}

The other important point of the present theory is that 
the ``unperturbed" term ${\cal L}_{0}$, i.e., $\hat{G}_{0}$, 
already contains the local correlation effect by $U_{\rm ff}$.
This effect plays a key role in critical phenomena in Ce- and Yb-based systems, 
which will be shown below to be the origin of the unconventional criticality. 
The local correlation effect emerges as dispersionless, almost $q$-independent 
$\chi^{\rm ffcc}_{0}({\bf q},0)$ and $\chi^{\rm cfcf}_{0}({\bf q},0)$ in Eq.~(\ref{eq:v2}), 
as shown in Fig.~\ref{fig:ffcc}(a). 
Here, the saddle point solution for $\exp(-S_{0})$ is employed 
for a typical parameter set of heavy-electron systems: 
$D=1$, $V=0.5$, and $U_{\rm ff}=\infty$ at total filling $n=7/8$ 
with $\varepsilon_{\bf k}={\bf k}^2/(2m_{0})-D$ and 
$n\equiv \bar{n}_{\rm f}+\bar{n}_{\rm c}$, 
where $\bar{n}_{\rm f}$ and $\bar{n}_{\rm c}$ are the number of f electrons and 
conduction electrons per ``spin" and site, respectively. 
The bare mass $m_{0}$ is chosen such that 
the integration from $-D$ to $D$ of the density of states of conduction electrons per ``spin" 
is equal to 1. 
This local nature is reflected in the inverse of valence susceptibility 
in Eq.~(\ref{eq:O2expand}) 
as an extremely small coefficient $A$. 
We note here that this flat-$q$ result is obtained not only for deep $\varepsilon_{\rm f}$ with 
$\bar{n}_{\rm f}=1/2$ in the Kondo regime, but also for shallow $\varepsilon_{\rm f}$ with 
$\bar{n}_{\rm f}<1/2$ in the valence-fluctuating regime (see Fig.~\ref{fig:ffcc}(b)). 
Here, we note that the c-f hybridization is always finite.

\begin{figure}
\includegraphics[width=70mm]{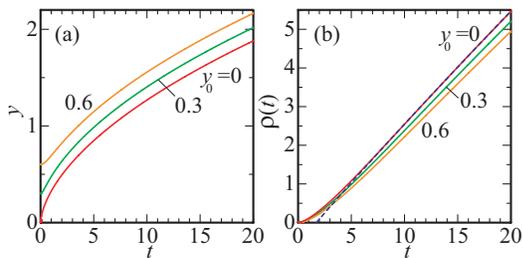} 
\caption{\label{fig:yt} (color online)
(a) Numerical solutions of Eq.~(\ref{eq:SCReq}) 
for $y_{0}=0.0$ (at QCP), 0.3, and 0.6 at $y_{1}=1$ and $x_{\rm c}=1$. 
(b) Electrical resistivity $\rho(T)$ calculated by using $y(t)$ in (a). 
Dashed line represents the linear-$t$ fit. 
}
\end{figure}

To clarify how this local nature causes unconventional criticality, we  
construct a self-consistent renormalization (SCR) theory for valance fluctuations.
Although higher order terms $v_{j}$ $(j\ge 3)$ in $S[\varphi]$ are irrelevant 
as shown in Eq.~(\ref{eq:vterms}), the effect of their mode couplings affects 
low-$T$ physical quantities significantly 
as is well known in spin-fluctuation theories~\cite{Moriya,MT,Hertz,Millis}. 
To construct the action using the best Gaussian taking account of the mode-coupling effects 
up to the 4-th order $(j\le 4)$ in $S[\varphi]$, we employ Feynman's inequality on the free energy: 
$
F\le F_{\rm eff}+T\langle S-S_{\rm eff}\rangle_{\rm eff}\equiv\tilde{F}(\eta)
$. 
Here, 
$
S_{\rm eff}[\varphi]=\frac{1}{2}\sum_{m}\sum_{{\bf q},{l}}
(\eta+Aq^2+C_{q}|\omega_{l}|)|\varphi_{m}({\bf q},i\omega_{l})|^{2}
$, 
and $\eta$ is determined to make $\tilde{F}(\eta)$ be optimum. 
By optimal condition $d\tilde{F}(\eta)/d\eta=0$, 
the self-consistent equation for $\eta$, i.e., the SCR equation, is obtained: 
$
\eta=\eta_{0}+3v_{4}\langle\varphi_{m}^{2}\rangle_{\rm eff}^{2}/N_{\rm s}
$, 
where 
$
\langle\varphi_{m}^{2}\rangle_{\rm eff}=T\sum_{{\bf q},l}(\eta+Aq^2+C_{q}|\omega_{l}|)^{-1}
$. 
Here, we write $\langle\varphi_{m}^{2}\rangle_{\rm eff}$ 
in a general form using $C_{q}$, 
which is given by $C_{q}\equiv C/\max\{q,l^{-1}_{\rm i}\}$ 
with $l_{\rm i}$ being the mean free path by impurity scattering~\cite{MNO}. 
When the system is clean, i.e., $C_{q}=C/q$, the SCR equation in $d=3$ 
in the $Aq_{\rm B}^{2}\lsim\eta$ regime with $q_{\rm B}$ being the momentum at the Brillouin Zone 
is given by 
\begin{eqnarray}
y=y_{0}+\frac{3}{2}y_{1}t\left[
\frac{x_{\rm c}^3}{6y}
-\frac{1}{2y}\int_{0}^{x_{\rm c}}dx\frac{x^3}{x+\frac{t}{6y}}
\right], 
\label{eq:SCReq}
\end{eqnarray}
where $y\equiv \eta/(Aq_{\rm B}^2)$, $t\equiv T/T_{0}$, $T_{0}\equiv Aq_{\rm B}^3/(2\pi C)$, 
$x\equiv q/q_{\rm B}$, $x_{\rm c}\equiv q_{\rm c}/q_{\rm B}$, 
and $y_{0}$ and $y_{1}$ are constants. 
When $y\gg t$, 
$y\propto t^{2/3}$ is obtained from Eq.~(\ref{eq:SCReq}) at the QCP with $y_{0}=0$.  
This indicates that the valence susceptibility shows unconventional criticality $\chi_{\rm v}({\bf 0},0)=\eta^{-1}\propto t^{-2/3}$. 
Figure~\ref{fig:yt}(a) shows numerical solutions of Eq.~(\ref{eq:SCReq}). 
Note here that the coefficient $A$ is quite small as shown above, giving rise to quite small $T_{0} (=Aq_{\rm B}^{3}/(2\pi C) \ll T_{\rm F})$ so that the region of $t\equiv T/T_{0}$ shown in Fig.~\ref{fig:yt} corresponds to that of $T\ll T_{\rm F}\sim O(D)$. 
Hence, a wide range of $t=T/T_{0}$ is shown in the plot.  
The least square fit of the data for $5\le t \le 100$ gives 
$y\propto t^{0.551}$. 
Since the Gaussian fixed point ensures the simultaneous divergence 
of valence susceptibility and uniform f-spin susceptibility as discussed above, 
$\chi^{\rm f}_{\rm s}$ shows divergent behavior $\sim t^{-\zeta}$ with $0.5 \lsim \zeta \lsim 0.7$ 
depending on the temperature range 
in agreement with experiments in YbRh$_2$(Si$_{0.95}$Ge$_{0.05}$)$_2$ and 
$\beta$-YbAlB$_4$. 
We note that the NMR or NQR relaxation rate is shown to be 
$(T_{1}T)^{-1}\sim \chi^{\rm f}_{\rm s}(t)\propto t^{-\zeta}$, which also quantitatively agrees with 
$(T_{1}T)^{-1}\sim T^{-0.5}$ in YbRh$_2$Si$_2$~\cite{Ishida}. 

We should note here that in the $T\to 0$ limit, although it may be experimentally difficult to access such a low temperature overcoming the smallness of $A$, the SCR equation follows the conventional $z=3$ type~\cite{Moriya,Hertz,Millis} in the clean system, giving rise to $y\propto t^{4/3}$. 
Then, at the QCP $(y_{0}=0)$, as $t$ decreases, a crossover from $y\propto t^{2/3}$ to $y\propto t^{4/3}$ occurs. 
In reality, however, because of the smallness of $A$, the low-$T$ range is extremely elongated by the relation $t=T/T_{0}$ with $T_{0}\propto A$, which makes it possible that 
unconventional criticality dominates over the experimentally accessible low temperature region. 

We note that the electrical resistivity $\rho(T)$ shows a $T$-linear dependence in the regime $t\gsim 5$ $(y\gsim 1)$ 
where Eq.~(\ref{eq:SCReq}) is applicable, as shown in Fig.~\ref{fig:yt}(b). 
Here, 
following a formalism of Ref.~\cite{Ueda1975}, 
$\rho(T)$ is calculated as 
$
\rho(T)\propto\frac{1}{T}\int_{-\infty}^{\infty}d\omega
\omega n(\omega)[n(\omega)+1]\int_{0}^{q_{\rm c}}dqq^3{\rm Im}\chi_{\rm v}^{\rm R}(q,\omega)
$
with $n(\omega)=1/({\rm e}^{\beta\omega}-1)$ being the Bose distribution function, and 
$
\chi_{\rm v}^{\rm R}(q,\omega)=(\eta+Aq^2-iC_{q}\omega)^{-1}
$, 
a retarded valence susceptibility. 
Here, $y(t)$ in Fig.~\ref{fig:yt}(a) is used for the clean system $C_{q}=C/q$, 
and the normalization constant is taken as 1 in the $\rho(t)$ plot. 
The emergence of $\rho(t)\propto t$ behavior can be understood from the locality of valence fluctuations:    
In the system with a small coefficient $A$, where the local character is strong, the dynamical exponent may be regarded as $z=\infty$ when we write $C_{q}$ in a general form as $C_{q}=C/q^{z-2}$. 
By using this expression in $\chi_{\rm v}^{\rm R}(q,\omega)$ in the calculation of $\rho(T)$ for $z=\infty$, we obtain $\rho(T)\propto T$ toward $T\to 0$~K. 
This result indicates that the locality of valence fluctuations causes the $T$-linear resistivity. 
The emergence of $\rho(T)\propto T$ by valence fluctuations was shown theoretically on the basis of the valence susceptibility $\chi_{\rm v}$ which has an approximated form for $z=\infty$ in Ref.~\cite{holms}. 
Our present formulation and the renormalization analysis provide a reasonable ground for the $\chi_{\rm v}$ introduced phenomenologically in Ref.~\cite{holms}. 

The evaluation of the quasiparticle self energy for a valence fluctuation exchange process by using the $\chi_{\rm v}$ shows that ${\rm Re}\Sigma(\varepsilon)\propto\varepsilon{\rm ln}(\varepsilon)$, which leads to  a logarithmic-$T$ dependence in the specific heat $C/T$  for a certain-$T$ range~\cite{WM}. The detailed $T$ dependence of $C/T$ will be discussed in a separated paper. 

In summary, we have shown that unconventional criticality commonly observed in YbRh$_2$Si$_2$, 
YbRh$_2$(Si$_{0.95}$Ge$_{0.05}$)$_2$, $\beta$-YbAlB$_4$, 
Ce$_{0.9-x}$Th$_{0.1}$La$_{x}$, and YbAuCu$_4$
is naturally explained from the viewpoint of the quantum valence criticality. 
The locality of the Yb or Ce valence fluctuation mode rising from local correlations of the 4f electrons 
is revealed to be the key origin. 
It is noted that our results may be regarded as an explicit manifestation of the marginal Fermi liquid theory~\cite{Varma1995}.


\end{document}